\documentclass[twocolumn,trackchanges]{aastex6}

\usepackage{amssymb,amsmath,amsthm}
\usepackage{graphicx}
\usepackage{mathrsfs}
\usepackage{booktabs}
\usepackage{multirow}
\usepackage{empheq}

\usepackage{natbib}
\shortauthors{Pesce et al.}

\begin{document}

\title{Submillimeter H$_2$O megamasers in NGC 4945 and the Circinus galaxy}

\author{D. W. Pesce}
\affil{Department of Astronomy, University of Virginia, 530 McCormick Road, Charlottesville, VA 22904, USA}
\email{dpesce@virginia.edu}
\and
\author{J. A. Braatz}
\affil{National Radio Astronomy Observatory, 520 Edgemont Road, Charlottesville, VA 22903, USA}
\and
\author{C. M. V. Impellizzeri}
\affil{National Radio Astronomy Observatory, 520 Edgemont Road, Charlottesville, VA 22903, USA}
\affil{Joint Alma Office, Alsonso de Cordova 3107, Vitacura, Santiago, Chile}

\begin{abstract}
We present 321 GHz observations of five AGN from ALMA Cycle 0 archival data: NGC 5793, NGC 1068, NGC 1386, NGC 4945, and the Circinus galaxy.  Submillimeter maser emission is detected for the first time towards NGC 4945, and we present a new analysis of the submillimeter maser system in Circinus.  None of the other three galaxies show maser emission, though we have detected and imaged the continuum from every galaxy.  Both NGC 4945 and Circinus are known to host strong ($\gtrsim 10$ Jy) 22 GHz megamaser emission, and VLBI observations have shown that the masers reside in the innermost $\sim 1$ parsec of the galaxies.  The peak flux densities of the 321 GHz masers in both systems are substantially weaker (by a factor of $\sim$100) than what is observed at 22 GHz, though the corresponding isotropic luminosities are more closely matched (within a factor of $\sim$10) between the two transitions.  We compare the submillimeter spectra presented here to the known 22 GHz spectra in both galaxies, and we argue that while both transitions originate from the gaseous environment near the AGN, not all sites are necessarily in common.  In Circinus, the spectral structure of the 321 GHz masers indicates that they may trace the accretion disk at radii interior to the 22 GHz masers.  The continuum emission in NGC 4945 and NGC 5793 shows a spatial distribution indicative of an origin in the galactic disks (likely thermal dust emission), while for the other three galaxies the emission is centrally concentrated and likely originates from the nucleus.
\end{abstract}


\section{Introduction}

Nuclear water vapor megamasers currently provide the only direct means to map gas in active galactic nuclei (AGN) on size scales of $\sim$0.1--1~pc.  Nearly all of the observational work on H$_2$O megamasers to date has focused on the $6_{16} - 5_{23}$ rotational transition at 22.235 GHz from the ortho-H$_2$O molecule \citep{2005ARA&A..43..625L}.  More than 160 galaxies have been detected in this line so far, the result of some $\sim$4000 galaxies surveyed (e.g., \citealt{2015IAUGA..2255730B}).  About 130 of the detections are associated with AGN, where they are called megamasers because of their large apparent luminosities.  The physical conditions that give rise to maser activity at 22 GHz are also compatible with masing in other transitions of the H$_2$O molecule, many of which fall in the submillimeter wavelength band (\citealt{1991ApJ...368..215N}; \citealt{2016MNRAS.456..374G}).

\cite{2005ApJ...634L.133H} presented the first observations of H$_2$O megamaser emission in a transition other than the 22 GHz, detecting maser emission at 183 GHz and (tentatively) at 439 GHz towards the galaxy NGC 3079.  This galaxy had previously been known to host strong 22 GHz masers \citep{1984A&A...141L...1H}, with VLBI observations confirming that the 22 GHz emission originates from the galactic nucleus (\citealt{1998ApJ...495..740T}, \citealt{2005ApJ...618..618K}).  Though the signal-to-noise of the (sub)millimeter detections ($\sim$7$\sigma$ for the 183 GHz transition) was too low to permit detailed study, the maser emission appears to arise from several narrow (spectrally unresolved) features spanning a velocity range comparable to that of the 22 GHz emission.

The 183 GHz transition was also detected towards Arp 220 by \cite{2006ApJ...646L..49C}, where it displays a very broad ($\sim$350~km~s$^{-1}$) and almost featureless spectral line structure.  Interestingly, this galaxy has not been detected in 22 GHz emission (e.g., \citealt{1986A&A...155..193H}), suggesting that the masing gas has a low density ($n_{\text{H}_2} \lesssim 10^6$~cm$^{-3}$) and temperature ($T \lesssim 100$~K).  From consideration of these physical conditions and the observed line width, \cite{2006ApJ...646L..49C} interpret the 183 GHz masers in this galaxy as likely originating from a large number ($\sim$10$^6$) of dense molecular cores rather than being associated with the galactic nuclei.

More recently, \cite{2013ApJ...768L..38H} used ALMA to detect 321 GHz H$_2$O megamaser emission towards the Circinus galaxy, another strong 22 GHz nuclear megamaser host (e.g., \citealt{2003ApJ...590..162G}).  The sensitivity of the Circinus observation was sufficient to showcase the richness of the high-frequency maser spectrum, opening up for the first time the possibility of using submillimeter masers in ways that had heretofore been restricted to the 22 GHz transition.

In this paper we report the first detection of submillimeter maser emission from NGC 4945, and we present a new calibration of the maser spectrum for the Circinus galaxy.  We note that \cite{2016arXiv160407937H} offer a parallel analysis of the NGC 4945 data presented here.  The observations and data reduction procedures are described in \S\ref{ObservationsAndReduction}, and in \S\ref{Discussion} we discuss the submillimeter emission and compare the 321 GHz masers to those at 22 GHz.  Throughout this paper we quote velocities using the optical definition in the heliocentric reference frame.

\section{Observations and data reduction} \label{ObservationsAndReduction}

We have analyzed archival Cycle 0 ALMA observations of five galaxies that are known to have strong (peak $S_{\nu} \gtrsim 200$~mJy) 22 GHz water maser emission associated with a central AGN: NGC 1068 \citep{1984Natur.310..298C}, NGC 1386 \citep{1996ApJS..106...51B}, NGC 4945 \citep{1979Natur.278...34D}, Circinus \citep{1982MNRAS.201P..13G}, and NGC 5793 \citep{1997PASJ...49..171H}.  All targets were observed at a rest-frame frequency of 321.226 GHz (ALMA Band 7), which corresponds to the $10_{2,9} - 9_{3,6}$ rotational transition of ortho-H$_2$O at an energy of $E_u/k \approx 1846$~K above ground\footnote{Frequencies, quantum numbers, and energy levels have been taken from Splatalogue: \url{http://www.cv.nrao.edu/php/splat/}.}.  NGC 5793 was further observed at a rest-frame frequency of 325.153 GHz, corresponding to the $5_{1,5} - 4_{2,2}$ rotational transition of para-H$_2$O at an energy of $E_u/k \approx 470$~K above ground.  The total bandwidth for each dual-polarization observation was 1.875~GHz, which was split into 3840 channels spaced contiguously every 0.488~MHz (corresponding to a velocity resolution of $\sim$0.5~km~s$^{-1}$).  The longest baselines for these observations were $\sim$360 meters (corresponding to a typical resolution of $\sim$$0.5''$), and there were between 18 and 25 antennas present (see Table \ref{tab:info}).

We obtained datasets and initial calibration scripts from the ALMA archive; all post-processing reduction, imaging, and spectral analysis was done using the Common Astronomy Software Applications package (CASA)\footnote{\url{http://casa.nrao.edu/}}.  Table \ref{tab:info} lists the observing parameters for each galaxy.

We detected and imaged continuum emission for all five sources (shown in Figure \ref{fig:Continuum_contours}), and in NGC 4945 the continuum was strong enough for self-calibration.  Two of the galaxies -- Circinus and NGC 4945 -- also host 321 GHz maser emission; we self-calibrated the Circinus data using the line emission.

\floattable
\begin{deluxetable}{lcccccc}
\tabletypesize{\scriptsize}
\tablecolumns{7}
\tablecaption{\label{tab:info}}
\tablehead{	&	\multicolumn{2}{c}{NGC 5793}	&	\colhead{Circinus}	&	\colhead{NGC 4945}	&	\colhead{NGC 1068}	&	\colhead{NGC 1386}}
\startdata
\multicolumn{1}{l|}{R.A. (J2000)}												&	\multicolumn{2}{c}{14:59:24.807}					&	14:13:09.906				&	13:05:27.279				&	02:42:40.770				&	03:36:46.237				\\
\multicolumn{1}{l|}{Dec. (J2000)}												&	\multicolumn{2}{c}{$-$16:41:36.55}				&	$-$65:20:20.468			&	$-$49:28:04.44			&	$-$00:00:47.84			&	$-$35:59:57.39			\\
\multicolumn{1}{l|}{$v_{\text{rec}}$ (km~s$^{-1}$)}			&	\multicolumn{2}{c}{3491}									&	434									&	563									&	1137								&	868									\\
\multicolumn{1}{l|}{Observing date (UTC)}								&	2012 Jun 01					&	2012 Jun 03					&	2012 Jun 03					&	2012 Jun 03					&	2012 Jun 06					&	2012 Aug 24					\\
\multicolumn{1}{l|}{$\nu_0$ (GHz)}											&	321.226							&	325.153							&	321.226							&	321.226							&	321.226							&	321.226							\\
\multicolumn{1}{l|}{$t_{\text{int}}$ (min.)}						&	6.3									&	21.0								&	19.1								&	15.3								&	15.8								&	11.6								\\
\multicolumn{1}{l|}{PWV (mm)}														&	1.35								&	0.40								&	0.55								&	0.60								&	0.54								&	0.64								\\
\multicolumn{1}{l|}{Antennas (number)}									&	21									&	20									&	18									&	18									&	20									&	25									\\
\multicolumn{1}{l|}{Flux calibrator}										&	Titan								&	Titan								&	Titan								&	Titan								&	Uranus							&	Uranus							\\
\multicolumn{1}{l|}{Bandpass calibrator}								&	3C 279							&	3C 279							&	3C 279							&	3C 279							&	3C 454.3						&	3C 454.3						\\
\multicolumn{1}{l|}{Phase reference}										&	J1517--243					&	J1517--243					&	J1329--5608					&	J1325--430					&	J0339--017					&	J0403--36						\\
\multicolumn{1}{l|}{Beam size ($''$)}										&	$0.55 \times 0.47$	&	$0.66 \times 0.46$	&	$0.66 \times 0.50$	&	$0.56 \times 0.52$	&	$0.66 \times 0.45$	&	$0.96 \times 0.53$	\\
\multicolumn{1}{l|}{Beam PA ($^{\circ}$)}								&	48									&	$-89$								&	$-18$								&	24									&	32									&	82									\\
\multicolumn{1}{l|}{RMS$_{\text{s}}$ (mJy)}							&	9.8									&	7.2									&	12.6$^{a}$					&	9.9									&	7.4									&	9.2									\\
\multicolumn{1}{l|}{RMS$_{\text{c}}$ (mJy beam$^{-1}$)}	&	0.39								&	0.29								&	0.48								&	3.0									&	0.42								&	0.36								\\
\midrule
\multicolumn{1}{l|}{$R_{\text{ap}}$ ($''$)}							&	2.5									&	2.5									&	2.0									&	5.0									&	1.5									&	1.0									\\
\multicolumn{1}{l|}{$S_{\nu}$ (mJy)}										&	10.8								&	18.8								&	90.8								&	733									&	47.1								&	4.3									\\
\multicolumn{1}{l|}{$\sigma_{S_{\nu}}$ (mJy)}						&	2.1									&	2.4									&	5.6									&	26.7								&	4.5									&	0.36$^{b}$					\\
\multicolumn{1}{l|}{$M_{\text{ISM}}$ (M$_{\odot}$)}			&	$4.0 \times 10^8$		&	$6.6 \times 10^8$		&	\ldots							&	$1.5 \times 10^8$		& \ldots							&	\ldots							\\
\enddata
\tablecomments{Information about the observations.  Listed coordinates (rows ``R.A." and ``Dec." for right ascension and declination, respectively) correspond to the tracking center entered for the observations, which might not precisely match the location of the target (we note in particular that the tracking center for NGC 4945 is displaced by approximately 2.5 arcseconds from the position listed in NED).  The ``$v_{\text{rec}}$" row lists the galaxy recession velocity in km~s$^{-1}$ (taken from NED), ``$\nu_0$" gives the rest-frame observing frequency, ``$t_{\text{int}}"$ denotes the on-source integration time in minutes, and ``PWV" is the average level of precipitable water vapor during the observation.  Half-power beam widths (``beam size" row) for the imaged data are given in arcseconds, and the beam position angles (``beam PA" row) are measured in degrees east of north.  The ``RMS$_s$" row lists the typical spectral sensitivity reached per 2~km~s$^{-1}$ vector-averaged channel, and the ``RMS$_c$" row gives the brightness sensitivity of the continuum image.  In general, the gradient in atmospheric opacity across a single spectrum causes the RMS$_s$ to increase by $\sim$30\% from one end to the other, so that the quoted value is an average.  The bottom section of the table lists the gas masses calculated from continuum observations.  $R_{\text{ap}}$ gives the radius of the aperture used to measure the continuum flux density (centered on the peak of the continuum emission), $S_{\nu}$ is the flux density measured inside of that aperture, $\sigma_{S_{\nu}}$ is the uncertainty in flux density, and $M_{\text{ISM}}$ is the ISM gas mass calculated using the method outlined in \S\ref{ContinuumEmission}. \\
$^a$The RMS$_s$ value for Circinus is given per 0.5~km~s$^{-1}$ channel.\\
$^b$Since the continuum emission in NGC 1386 is unresolved, we measure the peak flux density instead of the integrated, and we use the RMS of the continuum image as the uncertainty in this value.}
\end{deluxetable}

\subsection{Circinus} \label{CircinusReduction}

\begin{figure*}[t]
	\centering
		\includegraphics[width=0.75\textwidth]{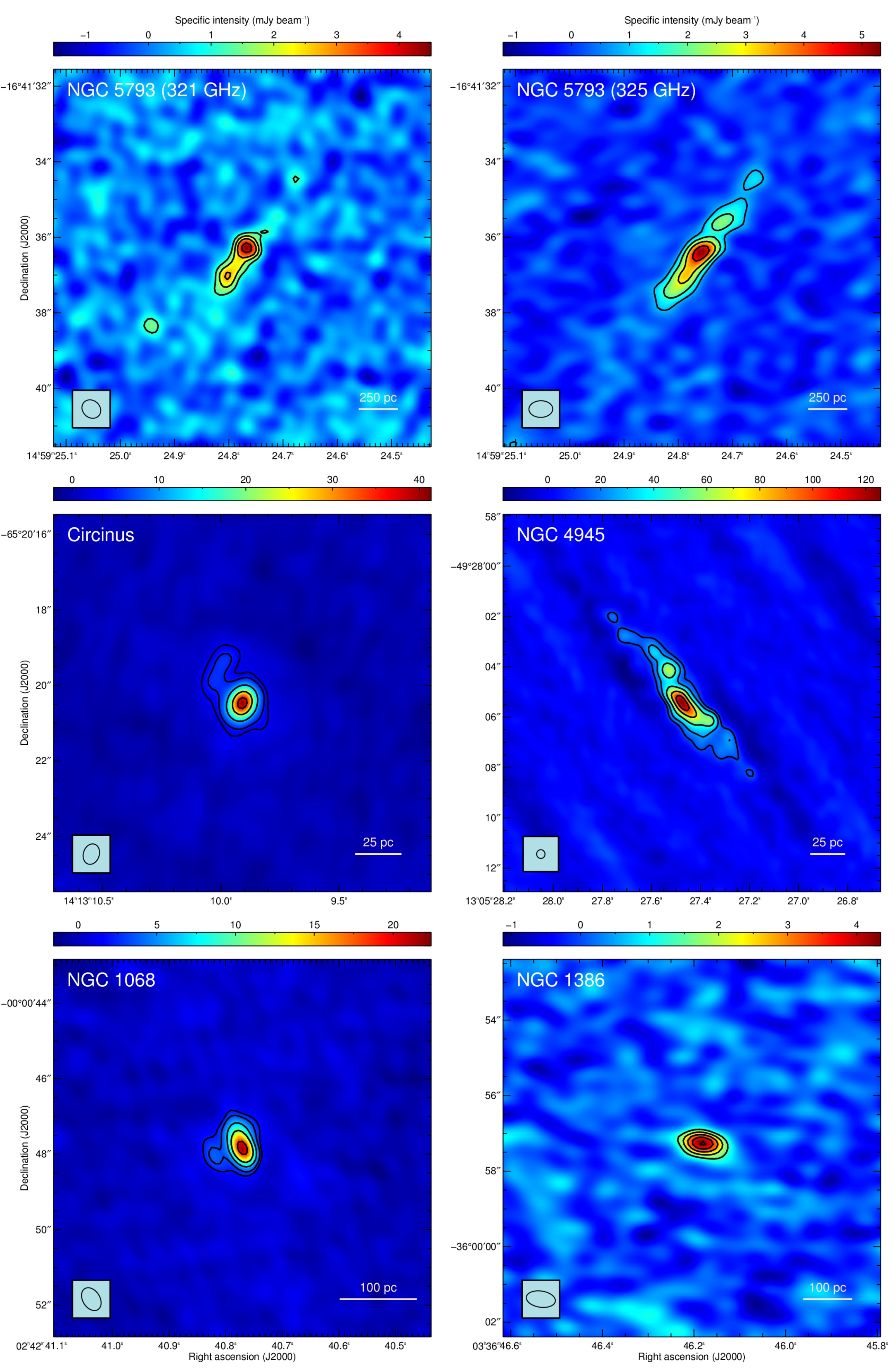}
	\caption{Continuum images.  \textsl{Top left}: 321 GHz image of NGC 5793, with $3\sigma$, $5\sigma$, $7\sigma$, and $10\sigma$ contours in black ($1 \sigma = 0.39$ mJy~beam$^{-1}$).  \textit{Top right}: 325 GHz image of NGC 5793, with $3\sigma$, $6\sigma$, $10\sigma$, $15 \sigma$, and $20\sigma$ contours in black ($1 \sigma = 0.29$ mJy~beam$^{-1}$).  \textit{Center left}: 321 GHz image of Circinus, with $5\sigma$, $10\sigma$, $25\sigma$, $50 \sigma$, and $75\sigma$ contours in black ($1 \sigma = 0.48$ mJy~beam$^{-1}$).  \textit{Center right}: 321 GHz image of NGC 4945, with $4\sigma$, $8\sigma$, $15\sigma$, $25 \sigma$, and $35\sigma$ contours in black ($1 \sigma = 3.0$ mJy~beam$^{-1}$).  \textit{Bottom left}: 321 GHz image of NGC 1068, with $5\sigma$, $8\sigma$, $15\sigma$, $25 \sigma$, and $45\sigma$ contours in black ($1 \sigma = 0.42$ mJy~beam$^{-1}$).  \textit{Bottom right}: 321 GHz image of NGC 1386, with $4\sigma$, $6\sigma$, $8\sigma$, $10 \sigma$, and $12\sigma$ contours in black ($1 \sigma = 0.36$ mJy~beam$^{-1}$).  Half-power restoring beam shapes are shown at the bottom left of each image, and scale bars are shown at the bottom right.  For NGC 5793, we adopt a Hubble law distance of 50~Mpc, using $H_0 = 70$~km~s$^{-1}$~Mpc$^{-1}$.  We use distances of 10.1~Mpc for NGC 1068 and 15.9~Mpc for NGC 1386; these were measured by \cite{2011A&A...532A.104N} and \cite{2013AJ....146...86T}, respectively, using the Tully-Fisher relation.}
	\label{fig:Continuum_contours}
\end{figure*}

\begin{figure*}[t]
	\centering
		\includegraphics[width=1.00\textwidth]{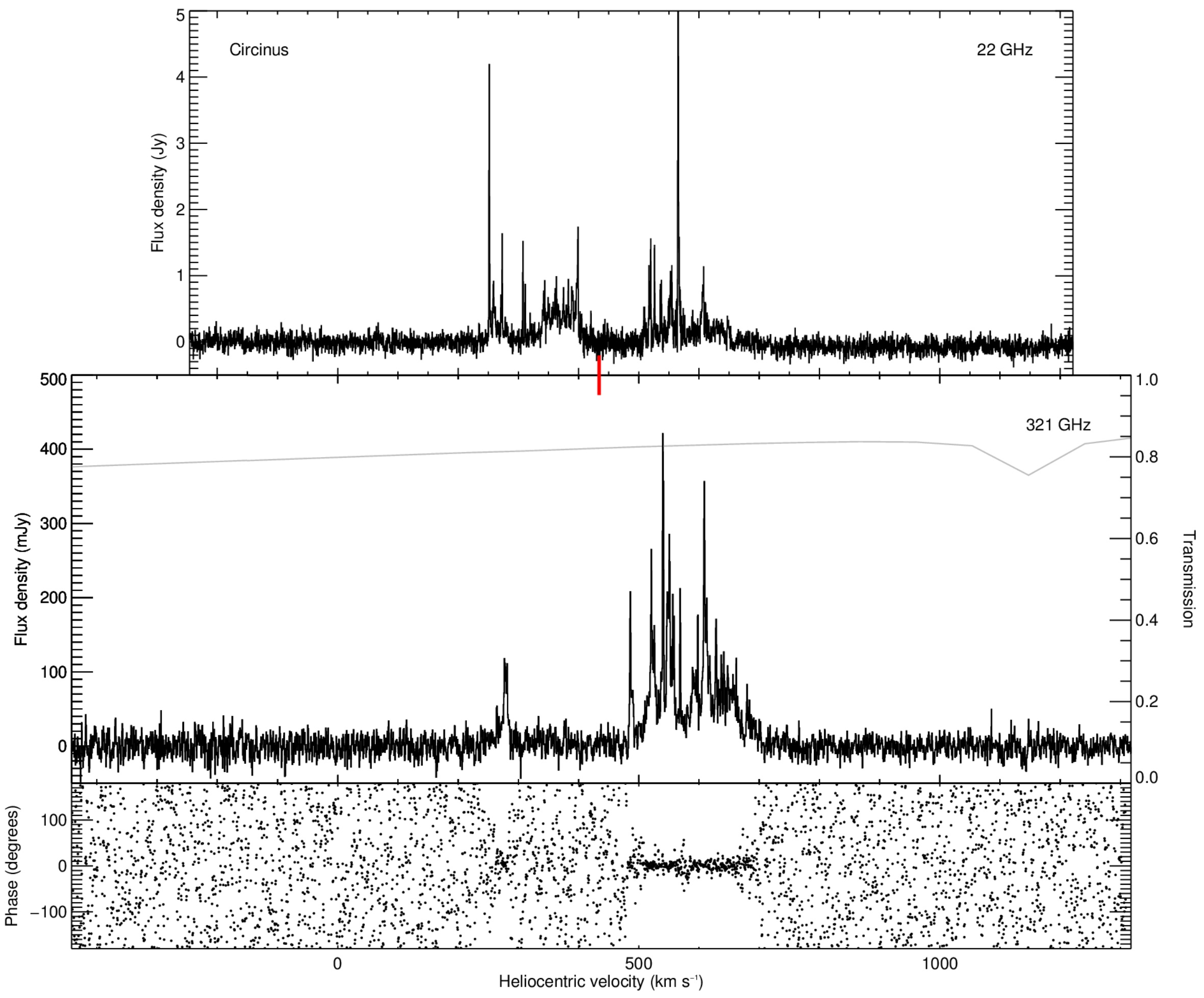}
	\caption{H$_2$O megamaser spectra of Circinus.  \textit{Top}: A reproduction of the 22 GHz spectrum taken with the 64~m Parkes telescope in 1998 August from \cite{2003ApJS..146..249B}.  We have restricted the vertical axis range to more easily see the weaker features, resulting in the strongest feature (peaking at $\sim$18~Jy) getting cut off.  \textit{Middle}: 321 GHz spectrum extracted from the continuum-subtracted data cube.  The channel width is 0.5~km~s$^{-1}$.  The atmospheric transmission curve, corresponding to a precipitable water vapor level matching that present during the observation, is overplotted in light gray.  Atmospheric transmission curves have been taken from the Atacama Pathfinder Experiment (APEX) transmission calculator (\url{http://www.apex-telescope.org/sites/chajnantor/atmosphere/}).  The recession velocity of the galaxy is marked by a vertical red line.  \textit{Bottom}: Phase plot for the calibrated 321 GHz spectrum.}
	\label{fig:Circinus_321GHz_spectra}
\end{figure*}

Initial imaging was performed using CASA task \texttt{clean} with natural UV weighting; after using \texttt{uvcontsub} (specifying line-free channels) to remove the continuum contribution, we separately imaged the line and continuum emission.  We then performed several iterations of phase-only self-calibration, using the $\sim$400 spectral channels with the strongest emission ($\gtrsim 100$ mJy, corresponding to the velocity range $\sim$500--700~km~s$^{-1}$) to determine the phase solutions.  We found that a solution interval of 1 minute (averaging both polarizations) was optimal, yielding sufficiently continuous solutions (i.e., consecutive phase solution jumps of $\lesssim 30^{\circ}$) to confidently interpolate the phases.  The calibration solutions were then applied to both the line and continuum data using \texttt{applycal}, and we stopped iterating self-calibration once there was no noticeable increase in signal-to-noise ratio (SNR).  We found that additional amplitude self-calibration did not improve the SNR, so we have retained the phase-only calibrations for analysis.  The resulting continuum image is shown in Figure \ref{fig:Continuum_contours}, and the spectrum extracted from the (spatially unresolved) line-only data cube is shown in Figure \ref{fig:Circinus_321GHz_spectra}.

\subsection{NGC 4945} \label{NGC4945Reduction}

The maser emission in NGC 4945 is not sufficiently strong for self-calibration, so we used the continuum emission instead.  Because the continuum emission in NGC 4945 is spatially resolved, we only used the longest baselines ($>150$~m, corresponding to the unresolved, point-like nuclear component of the emission) to determine the phase solutions that were then applied to the spectral line data; no such baseline restrictions were imposed when self-calibrating the continuum image itself.  We used a solution interval of 30 seconds, averaging both polarizations.  Despite repeated iterations of self-calibration, the sensitivity of the continuum image from this ``snapshot" observation remains dynamic-range limited (see \citealt{2011alma.book.....V}).  The resulting noise level of $3.0$~mJy~beam$^{-1}$ is thus larger than what one would nominally expect from a sensitivity calculation.

\begin{figure*}[t]
	\centering
		\includegraphics[width=1.00\textwidth]{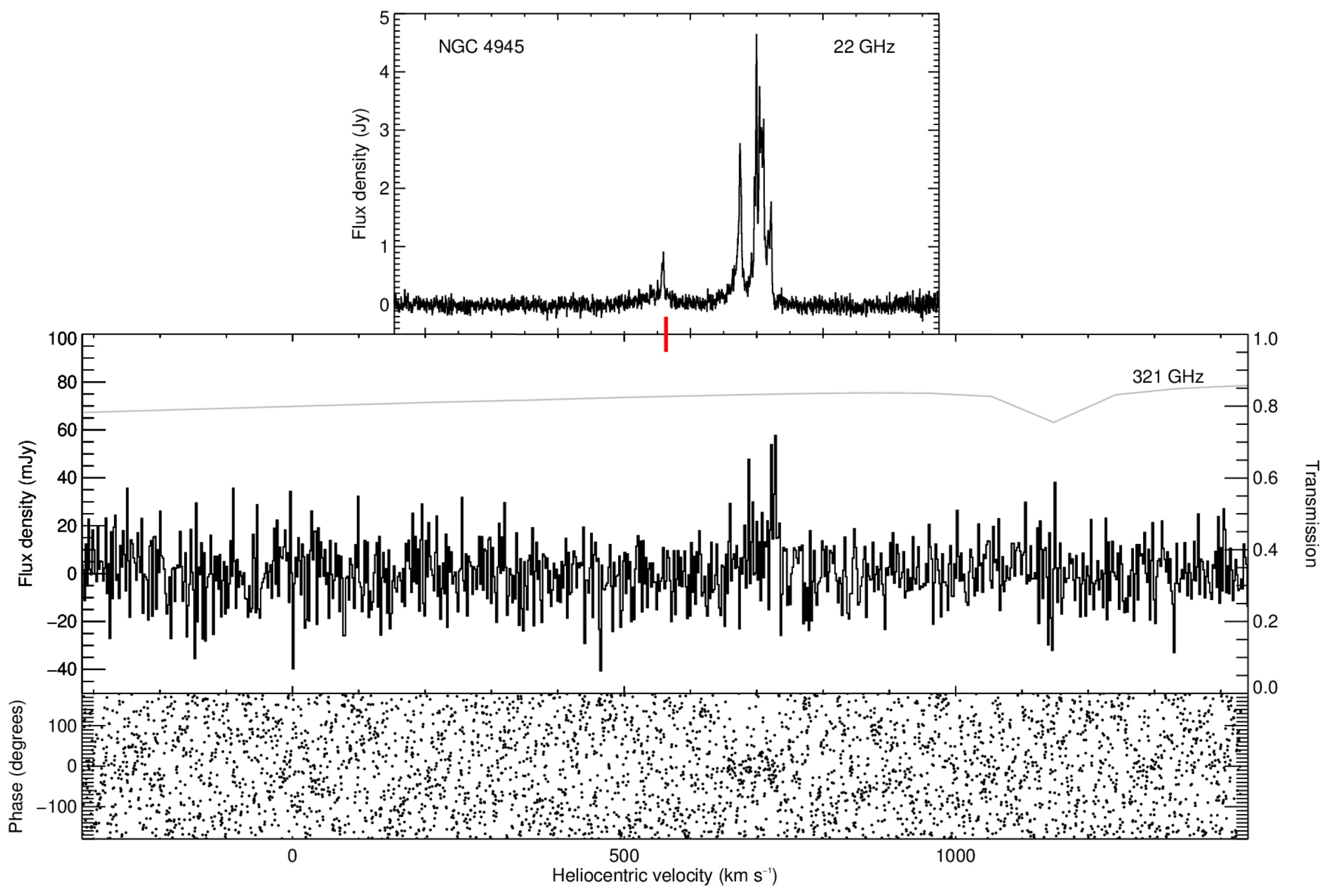}
	\caption{Same as Figure \ref{fig:Circinus_321GHz_spectra}, but for NGC 4945.  The channel size for the 321 GHz spectrum has been averaged to 2.0~km~s$^{-1}$.}
	\label{fig:NGC4945_321GHz_spectra}
\end{figure*}

The rest of the reduction procedure matches what was done for Circinus (see \S\ref{CircinusReduction}).  To account for the sizable ($\sim$$2.5''$) offset of the emission center from the phase center, we used \texttt{impbcor} to apply a primary beam correction prior to extracting a spectrum from the data cube.  The continuum image and spectrum for NGC 4945 are shown in Figures \ref{fig:Continuum_contours} and \ref{fig:NGC4945_321GHz_spectra}, respectively.

\section{Discussion} \label{Discussion}

\subsection{Continuum emission} \label{ContinuumEmission}

The continuum structures for NGC 5793 and NGC 4945 are both elongated in one direction (spanning $\sim$4$'' \approx 1000$~pc in NGC 5793, and $\sim$9$'' \approx 160$~pc in NGC 4945), and both appear to have substantial substructure.  Both of these galaxies are edge-on spirals, and the elongation axes of the submillimeter continua are aligned with the large-scale optical major axes (\citealt{1992MNRAS.258..296G}; \citealt{1997MNRAS.284..830E}).  The continuum in NGC 4945 is also resolved along the minor axis, spanning $\sim$1.5$'' \approx 30$~pc.  All of this indicates that the continuum emission in these galaxies traces the galactic disks, rather than originating from, e.g., a molecular torus region around the central AGN (though there may be a contribution to the emission in the centermost regions from such material).

At these wavelengths ($\lambda \approx 940$~$\mu$m), the continuum in NGC 5793 and NGC 4945 is likely dominated by optically thin thermal (i.e. blackbody) emission from large dust grains (see, e.g., \citealt{2003ARA&A..41..241D}; \citealt{2011A&A...525A.103C}).  The spectral energy distribution (SED) of such emission is typically modeled as a modified blackbody function (e.g., \citealt{2014A&A...571A..11P}), with the free parameters being the optical depth $\tau$, the dust temperature $T_d$, and the power-law index of the dust opacity $\beta$.  With only a single SED point per galaxy, we must assume fiducial values for two of these parameters (e.g., $\beta$ and $T_d$) to allow for a measurement of the third (e.g., $\tau$).  Further assumptions are then necessary to convert the optical depth to, e.g., a total interstellar medium (ISM) gas mass, $M_{\text{ISM}}$.

Fortunately, \cite{2014ApJ...783...84S} have developed an empirical calibration of the relationship between dust emission and ISM gas mass.  As long as the emission is measured in the Rayleigh-Jeans tail, the authors found that the calibration is relatively insensitive to whether the ISM is dominated by atomic or molecular gas, if the galaxy is normal or undergoing a starburst, or whether the dust lies in the inner or outer regions of the galaxy.  Rewritten in a suitable form, the conversion is given by

\begin{equation}
M_{\text{ISM}} = \frac{D^2 \lambda^2 S_{\nu}}{2 k \kappa_{\text{ISM}} T_d} . \label{eqn:DustConversion}
\end{equation}

Here, $D$ is the luminosity distance to the galaxy, $\lambda$ is the observing wavelength, $S_{\nu}$ is the observed flux density, $k$ is the Boltzmann constant, $\kappa_{\text{ISM}}$ is the dust opacity per unit mass of ISM, and $T_d$ is the dust temperature.  Most of the underlying physics here is contained in $\kappa_{\text{ISM}}$, which the authors calibrated using \textit{Planck} data to be

\begin{equation}
\left( \frac{\kappa_{\text{ISM}}}{4.84 \times 10^{-3} \text{ cm}^2 \text{ g}^{-1}} \right) = \left( \frac{\lambda}{850 \text{ $\mu$m}} \right)^{-\beta} . \label{eqn:ISMOpacity}
\end{equation}

When calculating gas masses, we use the results from \cite{2011A&A...536A..21P} to fix $\beta = 1.8$, and we adopt a dust temperature of $T_d = 25$~K (following \citealt{2014ApJ...783...84S}).  We measure the total continuum flux density for each galaxy using a circular aperture centered on the continuum peak, and we estimate the uncertainty using the dispersion of integrated flux densities measured in 15 non-overlapping, identical apertures that are offset from the continuum emission in the same image.  The results from these measurements are presented in the bottom portion of Table \ref{tab:info}.  The gas masses estimated from the 321~GHz and 325~GHz observations of NGC 5793 are broadly consistent, while the estimate for NGC 4945 is somewhat lower.

NGC 1068, NGC 1386, and Circinus all show continuum emission considerably more centrally-concentrated than in NGC 5793 and NGC 4945, so it is likely that AGN contributions to the continua for these galaxies are not negligible.  Disentangling the thermal (i.e., blackbody) and nonthermal (e.g., electron-scattered synchrotron, free-free) components of the emission is nontrivial, and requires multi-frequency observations (see, e.g., \citealt{2011ApJ...736...37K}).

For NGC 1068, we can compare our observations to those of \cite{2014A&A...567A.125G}, who used ALMA to map the continuum at 349~GHz down to a $1\sigma$ level of $0.14$~mJy~beam$^{-1}$.  Though this is a factor of $\sim$3 more sensitive than the map presented in this paper, we see consistent continuum structure and amplitude in the circumnuclear region (i.e., the region containing emission stronger than our sensitivity threshold) between the two observations.

\subsection{321 GHz H$_2$O masers in NGC 4945} \label{NGC4945Discussion}

The 321 GHz maser detection in NGC 4945 -- which represents the first time such emission has been seen in this galaxy -- is considerably fainter than in Circinus (Figure \ref{fig:NGC4945_321GHz_spectra}).  Individual maser features are detected at the $\sim$4--5$\sigma$ level, though the entire complex between 650 km~s$^{-1}$ and 750 km~s$^{-1}$ is detected at $\sim$$9 \sigma$ in integrated intensity.  We have calculated an isotropic luminosity using

\begin{equation}
L_{\text{iso}} = \frac{4 \pi D^2 \nu_0}{c} \int S_{v} dv . \label{eqn:IsotropicLuminosityOriginal}
\end{equation}

\noindent Here, $D$ is the distance to the galaxy, $\nu_0$ is the line rest frequency, and $S_{v}$ is the flux density as a function of velocity $v$.  Scaled to convenient units, this equation becomes

\begin{equation}
\left( \frac{L_{\text{iso}}}{L_{\odot}} \right) = 0.335 \left( \frac{D}{\text{Mpc}} \right)^2 \left( \frac{\int S_{v} dv}{\text{Jy~km~s$^{-1}$}} \right) . \label{eqn:IsotropicLuminosity}
\end{equation}

\noindent Adopting a distance to NGC 4945 of 3.7 Mpc \citep{2013AJ....146...86T}, the observed flux of $0.88$ Jy~km~s$^{-1}$ corresponds to an isotropic luminosity of $L_{\text{iso}} = 4$ L$_{\odot}$.  Though the flux density of individual features is down by a factor of $\sim$100 from what is observed at 22 GHz (e.g., \citealt{1996ApJS..106...51B}), the isotropic luminosity is only lower by a factor of $\sim$10.

Insofar as we are able to discern spectral structure, we see that it appears to match reasonably well with previous observations of NGC 4945 at 22 GHz (top panel of Figure \ref{fig:NGC4945_321GHz_spectra} has been reproduced from \citealt{2003ApJS..146..249B}).  The increasing feature strength with increasing velocity and the overall appearance of 2--3 dominant features are both reminiscent of the 22 GHz spectra.  However, the 321 GHz features at $\sim$687~km~s$^{-1}$ and $\sim$726~km~s$^{-1}$ (with possibly a third at $\sim$660~km~s$^{-1}$) don't map one-to-one with regions of 22 GHz emission.  Rather, and quite intriguingly, the 321 GHz peaks fall precisely where the 22 GHz emission drops off.

Unlike with Circinus, the origin of the 22 GHz emission from NGC 4945 is not yet well understood.  \cite{1997ApJ...481L..23G} made a VLBI map of NGC 4945 at 22 GHz using the southernmost antennas of the VLBA, and they found the spatial distribution of the masers to be approximately linear and distributed across $\sim$50~mas ($\sim$0.9~pc) from one end to the other.  This -- in particular the roughly symmetric location of redshifted and blueshifted emission to either side of the systemic velocity -- is suggestive of masers situated in an accretion disk.  The limited antennas available for mapping such a low declination source ($-49^{\circ}$) resulted in the map being rather incomplete (i.e., there were several systemic and blueshifted features that were too faint to map), but it is the best available for this source.  When measuring the positions of the 321 GHz maser spots, we found them to be spatially coincident (within the measurement uncertainties).  If the intrinsic distribution of the 321 GHz masers matches that of the 22 GHz masers, this is consistent with what we would expect for the $\sim$0.5$''$ beam and low signal-to-noise of the observations.

Working under the assumption that the 321 GHz emission traces material with the same kinematics as the 22 GHz, the observed 321 GHz features correspond only to the redshifted gas in the accretion disk.  If the 321 GHz spectral structure follows that of the 22 GHz emission, then the undetected blue and systemic features would be slightly below our detection threshold.  The low signal-to-noise in the current observations precludes any detailed characterization of this system, which must await higher sensitivity, better angular resolution observations than those presented here.

\subsection{321 GHz H$_2$O masers in Circinus} \label{CircinusDiscussion}

\cite{2013ApJ...768L..38H} discovered the 321 GHz maser in Circinus.  Here we re-examine the data, using strong maser lines to apply phase self-calibration (see \S\ref{CircinusReduction}).  The new calibration improves the SNR by a factor of $\sim$2 compared the initial analysis.

Published 22 GHz spectra of Circinus (e.g., top panel of Figure \ref{fig:Circinus_321GHz_spectra}, reproduced from \citealt{2003ApJS..146..249B}) show that the bulk of the maser emission occupies velocities between $\sim$250--650 km~s$^{-1}$ more or less contiguously, though often with a notable paucity of features near the systemic velocity.  \cite{2003ApJ...590..162G} (hereafter G03) observed Circinus between 1997 and 1998 using the Australia Telescope Long Baseline Array.  They detected two populations of masers, one arising from a warped accretion disk and the other associated with a wide-angle, bipolar outflow.

The 321 GHz masers are weaker in flux density by a factor of $\sim$30--100 compared to their 22 GHz counterparts.  Although the maser flux at 22 GHz is subject to interstellar scintillation \citep{1997ApJ...474L.103G}, this effect should be almost completely absent at 321 GHz (at such a high frequency, the diffractive scale of the turbulence will be much larger than the Fresnel scale; see \citealt{1992RSPTA.341..151N}).  At a distance to the galaxy of 4.2~Mpc (measured by \citealt{2013AJ....145..101K} using the Tully-Fisher relation), the observed flux of $17.5$~Jy~km~s$^{-1}$ corresponds to an isotropic luminosity (via Equation \ref{eqn:IsotropicLuminosity}) of $\sim$104~L$_{\odot}$; this is roughly a factor of four larger than the isotropic luminosity of the 22 GHz masers (e.g., \citealt{1996ApJS..106...51B}).

The 321 GHz and 22 GHz spectra share broad structural similarities.  Both have maser emission spanning comparable total velocity ranges and consolidated primarily into two groups located on either side of the systemic velocity, and in both cases the blueshifted group of features is weaker and sparser than the redshifted group.  We can also see that the region around the systemic velocity in the 321 GHz spectrum is devoid of obvious features -- either because no maser features exist at these velocities, or because they are below our detection threshold -- which is reminiscent of the same segment of the 22 GHz spectrum.

The VLBI maps from G03 show that the extent of the 22 GHz maser emission in Circinus is roughly $50 \times 80$~mas ($\sim$$1.0 \times 1.6$~pc), but as with NGC 4945 the 321 GHz maser spots are spatially co-located within our measurement uncertainties.  Though the absolute astrometric precision for ALMA observations is typically limited to $\sim$0.05 arcseconds without taking special calibration steps (\citealt{2015alma.book.....R}; \citealt{2014ARA&A..52..339R}), the relative uncertainty in point-source position within the same primary beam (as a fraction of the half-power beam width) is inversely proportional to the SNR (see, e.g., \citealt{1990A&AS...86..473F}).  Future high-resolution ALMA observations should thus have little difficulty mapping the 321 GHz masers in Circinus.

In lieu of a high angular resolution map, we can use the information contained in the spectrum to glean some understanding of the spatial distribution of the masers.  By applying a threshold proximity of 1 mas between any individual maser spot and the disk midline, G03 were able to assign a rough classification to each maser as originating from either the disk or the outflow.  In doing so, they found that the outflow masers dominated the emission between $\sim$300--600 km~s$^{-1}$, and that disk maser emission dominated blueward of $\sim$300~km~s$^{-1}$ and redward of $\sim$600~km~s$^{-1}$ (see Figure 6 in their paper).  With this picture from G03 as a guideline, we can compare the spectral distribution of the 22 GHz masers to that of the 321 GHz masers.

Their similar overall spectral structure suggests that the 321 GHz and 22 GHz masers are tracing roughly the same material.  This is to be expected from consideration of the physical conditions required for strong maser activity in these transitions.  \cite{2016MNRAS.456..374G} have performed a thorough exploration of the relevant parameter space (i.e., gas density, kinetic temperature, and dust temperature) and found that the 321 GHz transition shares an optimal gas density ($n_{\text{H}_2} \approx 10^9$ cm$^{-3}$) and collisional pumping scheme (i.e., low dust temperature) with the 22 GHz transition, though it prefers a somewhat larger kinetic temperature of $T_{K} \approx 1500$ K (compared to $T_K \approx 1000$ K for the 22 GHz transition)\footnote{We note that \cite{2016MNRAS.456..374G} modeled water maser emission in the context of evolved stars, and their models do not necessarily probe all of the conditions present in AGN central engines.  However, the calculations were performed assuming a minimally specific global geometry and dynamics (i.e., a plane-parallel medium with turbulence and a velocity gradient), and the explored region of parameter space covers the masing transitions relevant for this work (i.e., the 321 GHz and 325 GHz transitions).  We thus believe it to be suitable for the present level of analysis.}.  This could explain the apparent excess of 321 GHz maser emission between $\sim$650--750 km~s$^{-1}$, which is not typically seen in 22 GHz spectra (though we note that 22 GHz emission has been seen out to velocities as large as $\sim$900 km~s$^{-1}$, albeit with a much lower flux density than the bulk of the emission; see \citealt{2003ApJ...582L..11G}).  Under this interpretation, the 321 GHz emission redward of 650 km~s$^{-1}$ originates in the accretion disk at radii interior to where 22 GHz emission is found.

We see no features in the 321 GHz spectrum between $\sim$300--500 km~s$^{-1}$, which is a spectral range dominated by outflow emission at 22 GHz.  Either the 321 GHz emission does not trace the outflow at all or the 321 GHz outflow masers in this velocity range are much fainter than their 22 GHz counterparts (i.e., down by a larger factor from the higher-velocity emission to either side).  If the 321 GHz masers in fact only trace the disk emission, then the features detected between $\sim$500--600 km~s$^{-1}$ indicate that some of these masers must originate farther out in the disk than the 22 GHz masers.  Higher angular resolution observations will be able to discern whether any of the 321 GHz maser originate in the outflow or if they are all associated with the disk.

The possible high-velocity maser features -- seen at $\sim$1070~km~s$^{-1}$ and $\sim$1130~km~s$^{-1}$ in the 321 GHz spectrum -- coincide with an atmospheric line (see Figure \ref{fig:Circinus_321GHz_spectra}) and may represent elevated noise.

\section{Conclusions} \label{Conclusions}

We present 321 GHz ALMA observations of NGC 5793, NGC 1068, NGC 1386, NGC 4945, and the Circinus galaxy.  All galaxies are detected in continuum emission, and Circinus and NGC 4945 also display H$_2$O megamaser emission.  For NGC 4945 these data represent the first detection of submillimeter megamaser activity, while for Circinus we confirm the results of \cite{2013ApJ...768L..38H}, with an updated calibration.  In both cases the 321 GHz spectra appear structurally comparable to those of the 22 GHz masers.

The continuum emission in NGC 5793 and NGC 4945 is well-resolved and spatially extended along the optical major axes of these galaxies, which are both edge-on spirals.  This continuum is likely dominated by thermal emission from dust grains in the disk, and we use the observed fluxes to derive approximate ISM masses.  For the other three galaxies, the continuum emission is centrally-concentrated and thus likely contains a substantial non-thermal component from the AGN.

Though the 22 GHz maser emission in Circinus is associated with both the accretion disk and a molecular outflow, it is unclear whether the 321 GHz emission traces both environments or just the disk.  Furthermore, we have reason to believe that the 321 GHz masers trace the accretion disk at smaller radial separations from the central SMBH than the mapped 22 GHz masers do.  This can be confirmed by future ALMA observations of Circinus, which should seek to obtain a map of the maser features at the highest possible angular resolution.

\acknowledgments

We thank the anonymous referee for useful comments.  The National Radio Astronomy Observatory is a facility of the National Science Foundation operated under cooperative agreement by Associated Universities, Inc.  This paper makes use of the following ALMA data: 2011.0.00121.S.  ALMA is a partnership of ESO (representing its member states), NSF (USA) and NINS (Japan), together with NRC (Canada) and NSC and ASIAA (Taiwan) and KASI (Republic of Korea), in cooperation with the Republic of Chile. The Joint ALMA Observatory is operated by ESO, AUI/NRAO and NAOJ.  This research has made use of the NASA/IPAC Extragalactic Database (NED), which is operated by the Jet Propulsion Laboratory, California Institute of Technology, under contract with the National Aeronautics and Space Administration.

\facility{ALMA}
\software{CASA}


\end{document}